\documentclass[review,3p,12pt]{elsarticle}

\usepackage[export]{adjustbox}
\usepackage{hyperref}
\usepackage{graphicx}
\usepackage{amsmath}
\usepackage{mathtools}
\usepackage{esint}
\usepackage{amsfonts}
\usepackage{amssymb}
\usepackage{subfigure}
\usepackage{bm}
\usepackage{xcolor}
\usepackage{soul}
\usepackage{verbatim}
\usepackage{tikz}
\usepackage{tikz-3dplot}
\usepackage{dsfont}
\usetikzlibrary{arrows,shapes,bending}
\usepackage{pgfplots}
\usetikzlibrary{pgfplots.groupplots}
\usetikzlibrary{calc}
\usepackage{fp}
\usetikzlibrary{fixedpointarithmetic}   
\newlength\figureheight 
\newlength\figurewidth 
\usetikzlibrary{positioning}
\usetikzlibrary{matrix}
\usepackage{tikz-cd}
\usepackage{wrapfig}
\usepackage{accents}

\journal{International Journal of Solids and Structures}

\begin{document}

\begin{frontmatter}
	\title{A Geometric Formulation of  Linear Elasticity\\ Based on Discrete Exterior Calculus}
	
	\author{Pieter D.  Boom\corref{auc}}
	\ead{pieter.boom@manchester.ac.uk}
	\cortext[auc]{Corresponding author (+44-7547-975379)}
	
	\author{Odysseas Kosmas\fnref{au2}}
	\fntext[au2]{odysseas.kosmas@manchester.ac.uk}
	
	\author{Lee Margetts\fnref{au3}}
	\fntext[au3]{lee.margetts@manchester.ac.uk}
	
	\author{Andrey Jivkov\fnref{au4}}
	\fntext[au4]{andrey.jivkov@manchester.ac.uk}
	
	\address{Department of MACE, University of Manchester, George Begg Building, Manchester, M1 3BB, UK}

	\begin{abstract}
		A direct formulation of linear elasticity of cell complexes based on discrete exterior calculus is presented. The primary unknown are displacements, represented by primal vector-valued $0$-cochain. Displacement differences and internal forces are represented by primal vector-valued $1$-cochain and dual vector-valued $2$-cochain, respectively. The macroscopic constitutive relation is enforced at primal $0$-cells with the help of musical isomorphisms mapping cochains to smooth fields and vice versa. The balance of linear momentum is established at primal $0$-cells. The governing equations are solved as a Laplace equation with a non-local and non-diagonal material Hodge star. Numerical simulations of several classical problems with analytic solutions are presented to validate the formulation. Good agreement with known solutions is obtained. The formulation provides a method to calculate the relations between displacement differences and internal forces for any lattice structure, when the structure is required to follow a prescribed macroscopic elastic behaviour. This is also the first and critical step in developing formulations for dissipative processes in cell complexes.
	\end{abstract}


\end{frontmatter}



%
%

%
\section{Introduction}\label{sec:Intro}
%

Solids have discrete structures at a number of length scales, from atomic to polycrystalline, and their macroscopic behaviour emerges from finite rearrangements of these structures \cite{phillips:2001}. Analyses of discrete structures are performed by length-scale-specific mathematical formulations and corresponding numerical methods \cite{steinhauser:2017}. Examples include molecular dynamics at the atomic scale \cite{ward:2017}, particle-based methods, such as peridynamics \cite{silling:2000} and discrete elements \cite{zhu:2007}, and lattice-based methods \cite{wang:2008,phlipot:2019} at longer length scales. In many cases, the particle- and lattice-based methods are calibrated in such a way that the behaviour of the particle or lattice assembly matches a prescribed macroscopic (continuum) behaviour. Such calibrations are challenging and sometimes not possible for irregular arrangements of the assembly elements. However, the benefit of the discrete formulations is that existing, as well as emerging and evolving, material and geometric discontinuities are naturally captured and represented, which is beyond the capabilities of classical continuum formulations. In this respect, it will be beneficial to develop a generic mathematical formulation of deformation of finite discrete systems, from which the methods mentioned above can be either derived or calibrated. 

A natural starting point is the Discrete Exterior Calculus (DEC), which describes integration and differentiation through the topology and geometry of finite discrete cell complexes and their dual \cite{hirani:thesis,grady:2010}. DEC can be seen as a generalisation of the smooth exterior calculus, that satisfies the fundamental theorem of calculus (generalised Stoke's theorem) and Poincar\'e's Lemma. In DEC properties are intrinsically linked to the geometry of the cell components - vertices, edges, faces, volumes. For example, displacement are defined at vertices $[m/m^0]$, strain occurs along edges $[m/m^1]$, stress acts through faces $[N/m^2]$, and forces are in volumes $[N/m^3]$. A similar example for electromagnetism is presented in Gillette \cite{gillette:2009}. Physical interactions are defined by the topology of the complex, how the components are connected. For example, volumes interact through shared faces and conversely volumetric properties change as a result of the sum of fluxes through their bounding faces. Mathematically, the connections define exterior derivatives and enable common vector calculus operations to be mimicked, such as gradient, curl, divergence and Laplacian. More formal connections to vector fields can be made through the use of maps called musical isomorphisms \cite{abraham:1988}.

These properties make DEC an ideal platform to develop descriptions of physical processes which are fundamentally discrete. To date, DEC has been successfully applied to describe physical phenomena such as incompressible fluid flow \cite{mohamed:2016}, flow in porous media \cite{hirani:2015}, and electromagnetism \cite{chen:2016}. There have also been efforts to develop DEC formulations of elasticity. Yavari \cite{yavari:2008} outlined one such formulation, but without specifying how to prescribe the material constitutive relations. Angoshtari and Yavari \cite{angoshtari:2013} later presented a geometric description of incompressible linearised elasticity using DEC to identify a divergence-free solution space and recover the pressure field. However, strain is computed from the gradient of an interpolated displacement field, rather than using a discrete exterior derivative. DEC has also being used to inform finite-element approaches, for example Gillette and Bajaj \cite{gillette:2011}; however, this still relies on discretizing continuum descriptions of elasticity. To date no complete DEC formulation for elastic deformations has been proposed, a mandatory first step in extending DEC into modelling dissipative processes in discrete structures.

The aim of this work is to make this first step by presenting a geometric formulation of linear elasticity based on DEC. Brief reviews of elasticity and discrete exterior calculus are given in Sections \ref{sec:lin_elast} and \ref{sec:DEC}, respectively. The two are combined in Section \ref{sec:linDEC} to build a description of linear elasticity using DEC. The kinematics are defined on the primal complex and forces on the dual complex, similar to Yavari \cite{yavari:2008}, with the constitutive law applied through musical isomorphisms. Numerical results are presented for several classical problems in Section \ref{sec:numeric}. The development is discussed and conclusions drawn in Section \ref{sec:conc}.

%
\section{Elasticity}\label{sec:lin_elast}
%

The mechanics of an elastic body are defined by Newton's second law, providing equations for the balance of linear and angular momentum. The balance of linear momentum for a continuum body $\Omega$ can be written as
\begin{equation}\label{eq:governing:tensor}
	\nabla\cdot\tilde{\tau}(\tilde{x}) + \tilde{f}(\tilde{x}) = \tilde{\rho}\tilde{a}(\tilde{x}), \quad\text{for }\tilde{x}\in\Omega, 
\end{equation}
where $\tilde{\tau}$ is a stress tensor\footnote{In this article vector fields are marked with tildes $\tilde{\square}$ to distinguish them from discrete quantities.}\footnote{Solid mechanics and discrete exterior calculus commonly use symbol $\sigma$ to represent stress and simplicies, respectively. Therefore, in this article the stress tensor field will be denoted $\tilde{\tau}$ and ${\sigma}$ used for simplicies.}, $\tilde{f}$ are volumetric body forces, and $\tilde{\rho}\tilde{a}$ are the forces due to acceleration. In this article we will be concerned with static problems, where the inertial forces are zero, $\tilde{\rho}\tilde{a}=0$.

The deformation of the the body is defined as a map between a reference configuration $\tilde{X}$ and the deformed configuration $\tilde{x}=\tilde{X} + \tilde{u}$, where $\tilde{u}$ is the displacement field. Stress can be derived from the deformation through a constitutive law, such as Hooke's law or the Mooney-Rivlin model for neo-Hookean solids. 

The simplest relationship is for a linear elastic isotropic material. In this case the strain tensor, under the infinitesimal strain assumption, reduces to
\begin{equation}\label{eq:strain}
	\tilde{\epsilon} = \frac{1}{2}\left(\nabla \tilde{u} + (\nabla \tilde{u})^T\right),
\end{equation}
%
and stress is defined through the linear constitutive relationship
\begin{equation}\label{eq:constitutive}
	\tilde{\tau} = \lambda\ \text{tr}(\tilde{\epsilon})\tilde{I}+ 2\mu\tilde{\epsilon} ,
\end{equation}
where $\tilde{I}$ is the identity tensor, $\lambda$ is Lam\'e's first constant, and  $\mu$ is the shear modulus. We assume in this article that these constants do not vary as a function of either spatial location or time. The symmetry of the stress tensor satisfies the continuum requirement for balance of angular momentum.

The resulting governing equation for the balance of linear momentum becomes well-posed with the addition of essential (Dirichlet) and natural (Neumann) conditions on the boundary $\partial\Omega$ of the body. Traditionally, the natural conditions take the form
\begin{equation}\label{eq:stressBC}
	\tilde{\tau}\cdot\tilde{n}_i = \tilde{t}_i \in\partial\Omega_i
\end{equation}
where $\tilde{n}_i$ is the outward pointing unit normal of the boundary surface patch $\partial\Omega_i$, and $\tilde{t}_i$ is prescribed surface traction.

%
\section{Discrete exterior calculus}\label{sec:DEC}
%

This section is intended to give a brief introduction to discrete exterior calculus (DEC). The theory and notation presented are based on Refs. \cite{desbrun:2005,hirani:thesis,grady:2010}. Readers familiar with DEC can proceed to the next section, where DEC is applied to the equations of linear elasticity.

Calculus describes change through integration and differentiation. The familiar vector calculus is concerned with the behaviour of fields in three-dimensional Euclidean space. Smooth exterior calculus generalises vector calculus to differential manifolds in arbitrary dimensions \cite{abraham:1988}. Likewise, discrete exterior calculus is an extension to oriented piecewise flat manifolds \cite{grady:2010}. The specific formulation used presently is based on simplicial complexes in arbitrary dimensions \cite{hirani:thesis}. DEC satisfies critical identities such as Stoke's theorem and Poincar\'e's Lemma, and provides analogues to the gradient, divergence, and curl.

\subsection{Primal simplicial complex}

A solid body $\Omega$ is represented as an oriented three-dimensional simplicial complex $K$. An example for a circular cylinder is shown in Figure \ref{fig:complex}. A simplicial complex is composed of $k$-dimensional simplices ${\sigma}^k=\{v_0,\ldots,v_k\}$, where the $v_i$ identify the vertices of each simplex, and the order of the elements $v_i$ defines an orientation. The boundary of each $k$-simplex is formed by $(k-1)$-simplices whose orientations are imposed by the parent simplex. In DEC, the orientations of top-dimension $k$-simplices are chosen such that the $(k-1)$-simplices at interfaces have opposite orientations. All of these conventions are shown pictorially in Figure \ref{fig:simplex}.

\begin{figure}[!t]
\centering
	\includegraphics[width=0.35\textwidth]{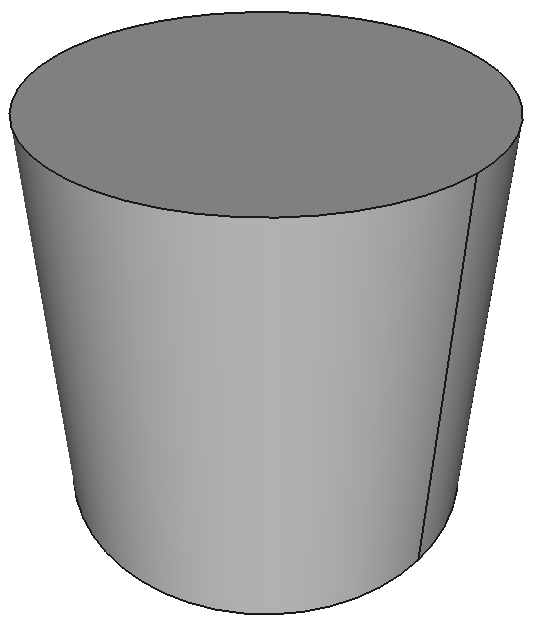}\hspace{1cm}
	\includegraphics[width=0.35\textwidth]{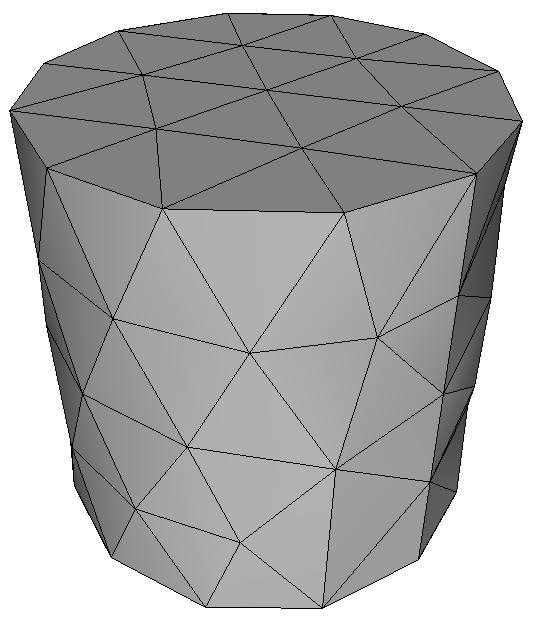}
	\caption{Cylinder and corresponding primal simplicial complex\label{fig:complex}}
\end{figure}
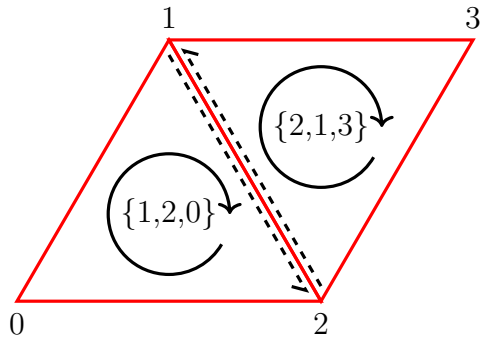
\begin{figure}[!t]
\centering
	\begin{tikzpicture}[scale=4]
		
	\draw[very thick, red] (0,0) -- (0.5,0.86602540378) -- (1,0) --cycle;
	\draw[very thick, red] (1,0) -- (0.5,0.86602540378) -- (1.5,0.86602540378) --cycle;
	
	\draw(0,-0.075) node{0};
	\draw(0.5,0.86602540378+0.075) node{1};
	\draw(1,-0.075) node{2};
	\draw(1.5,0.86602540378+0.075) node{3};
	
	\draw(0.5,1/3*0.86602540378) node{\{1,2,0\}};
	\draw(1,2/3*0.86602540378) node{\{2,1,3\}};
	
	\draw[very thick,<-] (0.7,1/3*0.86602540378) arc (0:330:0.2);
	\draw[very thick,<-] (1.2,2/3*0.86602540378) arc (0:330:0.2);
	
	\draw[very thick, dashed,-{Straight Barb[right]}] (1.0,0.05) -- (0.5+0.5*0.086602540378,0.86602540378-0.025);
	\draw[very thick, dashed,-{Straight Barb[right]}] (0.5,0.86602540378-0.05) -- (1-0.5*0.086602540378,0.025);
	\end{tikzpicture}
	\caption{Two-dimensional simplices shown with vertex numbering, orientation, and imposed orientation on the interfacing edge.\label{fig:simplex}}
\end{figure}

\subsection{Chains and cochains}
One of the primary operations in calculus is integration. This requires a domain of integration, which can be a subset of the entire domain in which a function is defined. A $k$-chain assigns an integer value to each $k$-simplex in a given complex $K$ and can be interpreted as an oriented indicator for the domain of integration. A value of zero indicates a particular cell is not included in the domain of integration; a negative value indicates an orientation opposite to a $k$-simplex's reference orientation; and values greater than one indicate multiplicity in a set. Similarly, a $k$-cochain assigns a scalar value to each $k$-simplex in a complex $K$, but is interpreted as a linear functional, locally mapping $k$-chains to scalars at each $k$-simplex. Thus, integration can be viewed as the product of a $k$-chain and $k$-cochain.

\subsection{Discrete exterior derivatives}
The other primary operation is differentiation. The relationship between integration and differentiation is defined by Stokes theorem, which relates the value of a function integrated over the boundary of an oriented manifold to the value of its differential integrated over the entire manifold. In DEC, the differential operator is called the discrete exterior derivative: a map from $(k-1)$-cochains to $k$-cochains. Recall that the boundary of a $k$-simplex is formed by $(k-1)$-simplices, and note the parallel to Stokes theorem. Formally, a discrete exterior derivative is the oriented map
\begin{equation}
	d^{k} : C^{k-1} \rightarrow C^{k},
\end{equation}
where $C^k$ is the space of all $k$-chains in a complex $K$. These maps are signed adjacency matrices:
\begin{equation*}
d^{k}_{ij} = 
     \begin{cases}
       \ 0 & \text{if} \quad \sigma_i^{k-1}\quad\text{is not on the boundary of }{\sigma}_j^{k}\\
       \ 1 & \text{if} \quad \sigma_i^{k-1}\quad\text{is on the boundary of }{\sigma}_j^{k} \text{ with coherent orientation}\\
       -1 & \text{if} \quad \sigma_i^{k-1}\quad\text{is on the boundary of }{\sigma}_j^{k} \text{ with incoherent orientation}
     \end{cases}.
\end{equation*}
The requirement that top-dimension simplices have consistent orientations ensures that Stokes theorem holds for a domain of integration that contains an arbitrary number of simplices in a complex. In other words, the contribution from all interfacial $(k-1)$-simplices sums to zero. The construction of the discrete exterior derivative also satisfies Poincare's Lemma: the boundary of a boundary is empty, $d^{k}d^{k-1} = 0$. Note , that the discrete exterior derivatives do not involve the metric of the space. In other words, they are purely topological, or combinatorial, rather than geometric, and therefore do not change the unit/dimension of the differentiated cochain.

\subsection{Dual cell complex}
\begin{figure}[!t]
\centering
	\begin{tikzpicture}[scale=3]
	
	\draw[very thick, red] (0,0) -- (0.5,0.86602540378) -- (1,0) --cycle;
	\draw[very thick, red] (0,0) -- (0.5,-0.86602540378) -- (1,0) --cycle;
	\draw[very thick, red] (1,0) -- (0.5,0.86602540378) -- (1.5,0.86602540378) --cycle;
	\draw[very thick, red] (1,0) -- (0.5,-0.86602540378) -- (1.5,-0.86602540378) --cycle;
	\draw[very thick, red] (2,0) -- (1.5,0.86602540378) -- (1,0) --cycle;
	\draw[very thick, red] (2,0) -- (1.5,-0.86602540378) -- (1,0) --cycle;
	
	\draw[draw=blue,thick] (0.25,0.5*0.86602540378) -- (0.5,1/3*0.86602540378)  -- (0.5,0) ;
	\draw[draw=blue,thick] (0.25,0.5*0.86602540378) -- (0.5,1/3*0.86602540378)  -- (0.75,0.5*0.86602540378);
	\draw[draw=blue,thick] (0.75,0.5*0.86602540378) -- (0.5,1/3*0.86602540378) -- (0.5,0);
	
	\draw[draw=blue,thick] (1.25,0.5*0.86602540378) -- (1.5,1/3*0.86602540378)  -- (1.5,0);
	\draw[draw=blue,thick] (1.25,0.5*0.86602540378) -- (1.5,1/3*0.86602540378)  -- (1.75,0.5*0.86602540378);
	\draw[draw=blue,thick] (1.75,0.5*0.86602540378) -- (1.5,1/3*0.86602540378) -- (1.5,0);
	
	\draw[draw=blue,thick] (0.75,0.5*0.86602540378) -- (1,2/3*0.86602540378)  -- (1.25,0.5*0.86602540378);
	\draw[draw=blue,thick] (0.75,0.5*0.86602540378) -- (1,2/3*0.86602540378)  -- (1,0.86602540378);
	\draw[draw=blue,thick] (1.25,0.5*0.86602540378) -- (1,2/3*0.86602540378)  -- (1,0.86602540378);
	
	\draw[draw=blue,thick] (0.75,0.5*-0.86602540378) -- (1,2/3*-0.86602540378)  -- (1.25,0.5*-0.86602540378);
	\draw[draw=blue,thick] (0.75,0.5*-0.86602540378) -- (1,2/3*-0.86602540378)  -- (1,-0.86602540378);
	\draw[draw=blue,thick] (1.25,0.5*-0.86602540378) -- (1,2/3*-0.86602540378)  -- (1,-0.86602540378);
	
	\draw[draw=blue,thick] (0.25,0.5*-0.86602540378) -- (0.5,1/3*-0.86602540378)  -- (0.5,0);
	\draw[draw=blue,thick] (0.25,0.5*-0.86602540378) -- (0.5,1/3*-0.86602540378)  -- (0.75,0.5*-0.86602540378);
	\draw[draw=blue,thick] (0.75,0.5*-0.86602540378) -- (0.5,1/3*-0.86602540378) -- (0.5,0);
	
	\draw[draw=blue,thick] (1.25,0.5*-0.86602540378) -- (1.5,1/3*-0.86602540378)  -- (1.5,0);
	\draw[draw=blue,thick] (1.25,0.5*-0.86602540378) -- (1.5,1/3*-0.86602540378)  -- (1.75,0.5*-0.86602540378);
	\draw[draw=blue,thick] (1.75,0.5*-0.86602540378) -- (1.5,1/3*-0.86602540378) -- (1.5,0);
	\end{tikzpicture}
	\caption{Two-dimensional primal simplical complex (red) and dual cell complex (blue). \label{fig:dual}}
\end{figure}
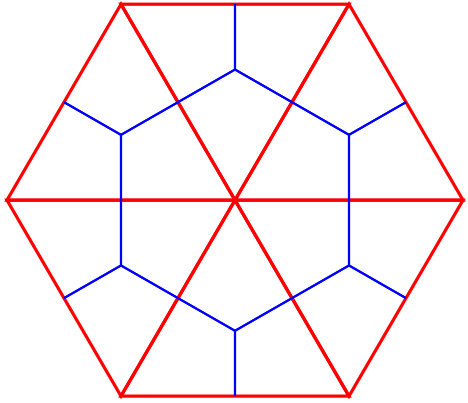

The discrete exterior derivatives provide useful maps from cochains on low-dimension simplices to those on higher-dimension simplices, but not vise-versa. To create meaningful maps in the other direction, a second three-dimensional oriented cell complex $\star K$ is constructed which is dual to the primal simplicial complex. The dual cell complex is constructed such that each $k$-dimensional cell, or $k$-cell ${\sigma}^{\star k}$, is associated one-to-one with a $(3-k)$-simplex in the primal, including it's orientation. An example of associated primal and dual complexes is shown in Figure \ref{fig:dual}. The geometric ordering of the dual-cells is in the opposite direction to the primal-simplices (See Figure \ref{fig:derham}). Therefore, discrete exterior derivatives on the dual $d^{\star k}$ will map in the opposite direction to those on the primal. Conveniently, given the relationship between the primal and dual, the discrete exterior derivatives for the dual complex are the transpose of those for the primal $d^{\star k} = (d^{(2-k)})^T$. 

Note, that the dual complex is uniquely defined combinatorially, but not geometrically. The geometric construction of the dual depends on a selection of the dual $0$-cells. If the dual $0$-cells are selected to be the circumcentres of the primal $3$-cells, the resulting dual complex is the Voronoi tessellation around the primal $0$-cells; the primal $2$-cells are orthogonal to the dual $1$-cells, and the primal $1$-cells are orthogonal to the dual $2$-cells. Clearly, any other selection is possible, including the occasionally used barycentric dual, i.e. dual $0$-cells are the barycentres of the primal $3$-cells. This work will use the circumcentric (Voronoi) dual.

\subsection{Discrete Hodge star}
To make use of the discrete exterior derivatives on the dual complex, further maps are required between cochains on the primal and dual complexes. These isomorphism  are called discrete Hodge stars $\star^k:C^k\rightarrow D^{3-k}$, where $D^k$ is the space of all $k$-chains in the dual complex $\star K$. In contrast to the discrete exterior derivatives, the discrete Hodge stars do involve the metric of the space. When the dual complex is the Voronoi dual, the discrete Hodge stars are diagonal maps of the form:

\begin{equation}
	\star^k_{ii} = \frac{|{\sigma}_i^{\star(3-k)}|}{|{\sigma}_i^k|},
\end{equation}
where $|\square|$ denotes the volume of the simplex or cell, with the convention that vertices are dimensionless and have unit volume. Thus, the Hodge stars have the following dimensions:
\begin{equation}
	\star^0 = [m^3],\quad\star^1=[m^1],\quad\star^2=[m^{-1}],\quad\star^3=[m^{-3}].
\end{equation}

\subsection{de Rham complex}
\begin{figure}[!t]
\centering
\hspace{-0.3cm}
\includegraphics[valign=m,width=0.25cm]{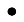} \hspace{1.75cm}
\includegraphics[valign=m,width=1.5cm]{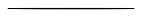} \hspace{1.5cm}
\includegraphics[valign=m,width=1cm]{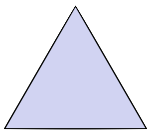} \hspace{1.5cm}
\includegraphics[valign=m,width=1cm]{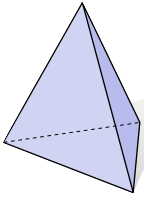} \\
\begin{tikzpicture}[scale=1.5]
\draw (0,0) node {$C^0$};
\draw (1,0.25) node {$d^0$};
\draw[thick, ->] (0.5,0) -- (1.5,0);
\draw (2,0) node {$C^1$};
\draw (3,0.25) node {$d^1$};
\draw[thick, ->] (2.5,0) -- (3.5,0);
\draw (4,0) node {$C^2$};
\draw (5,0.25) node {$d^2$};
\draw[thick, ->] (4.5,0) -- (5.5,0);
\draw (6,0) node {$C^3$};

\draw[thick, ->] (0,-0.5) -- (0,-1.5);
\draw (0.25,-1) node {$\star^0$};
\draw (0,-2) node {$D^3$};
\draw (1,-2+0.25) node {$d^{\star2}$};
\draw (1,-2-0.25) node {$(d^{0})^T$};
\draw[thick, <-] (0.5,-2) -- (1.5,-2);
\draw[thick, ->] (2,-0.5) -- (2,-1.5);
\draw (2+0.25,-1) node {$\star^1$};
\draw (2,-2) node {$D^2$};
\draw (3,-2+0.25) node {$d^{\star1}$};
\draw (3,-2-0.25) node {$(d^{1})^T$};
\draw[thick, <-] (2.5,-2) -- (3.5,-2);
\draw[thick, ->] (4,-0.5) -- (4,-1.5);
\draw (4+0.25,-1) node {$\star^2$};
\draw (4,-2) node {$D^1$};
\draw (5,-2+0.25) node {$d^{\star0}$};
\draw (5,-2-0.25) node {$(d^{2})^T$};
\draw[thick, <-] (4.5,-2) -- (5.5,-2);
\draw[thick, ->] (6,-0.5) -- (6,-1.5);
\draw (6+0.25,-1) node {$\star^3$};
\draw (6,-2) node {$D^0$};
\end{tikzpicture}
\ \\
\hspace{-0.75cm}
\includegraphics[valign=m,width=1cm]{tet} \hspace{1.65cm}
\includegraphics[valign=m,width=1cm]{tri} \hspace{1.5cm}
\includegraphics[valign=m,width=1.5cm]{line} \hspace{1.75cm}
\includegraphics[valign=m,width=0.25cm]{node} 
\caption{de Rham complex for three dimensional primal and dual complexes \label{fig:derham}}
\end{figure}

The maps between the simplices and cells in the primal and dual complexes are summarized in the de Rham complex shown in Figure \ref{fig:derham}.

\subsection{Musical isomorphisms}

In applications, it is necessary to map between fields (scalar, vector, tensor, etc) and cochains, or vise versa. This may be to set initial or boundary conditions or apply certain pointwise relationships defined traditionally for continuous formulations. These maps are called musical isomorphisms and involve the metric of the space. 

\subsubsection{Discrete flat musical isomorphism}
Discrete flat $\flat$ musical isormorphisms are maps from vector fields to cochains. They mimic directional integrals of vector fields onto piecewise smooth manifolds, the simplices and cells of the primal and dual complexes in this case. The discrete flats to $1$-cochains presented here are the unique flats that yield a valid discrete divergence theorem, though several others have been proposed \cite{hirani:thesis}. The discrete flats used in this article are summarized below for a vector field $\tilde{u} \in \mathbb{R}^3$:
\begin{equation}
\begin{array}{l}
\flat^{0}\tilde{u} = {\mathtt{u}}^{0}_{i}
	=\vec{{\sigma}}^{0}_{i} \tilde{u}\bigg|_{{\sigma}^{0}_{i}}
,\\[2ex]
\flat^{1}\tilde{u} = {\mathtt{u}}^{1}_{i}
	=\left(\sum_{\forall \sigma^{0}_{j}\prec{\sigma}^{1}_{i}}\left|{\sigma}^{1}_{i} \cap \sigma^{\star 3}_{j}\right|\, \tilde{u}\bigg|_{{\sigma}^{0}_{j}}\right)\cdot\vec{{\sigma}}^{1}_{i}
	\approx \int_{{{\sigma}}^{1}_{i}} \tilde{u}\cdot d\vec{{\sigma}}^{1}_{i}
, \\[2ex]
\flat^{\star0}\tilde{u}   = {\mathtt{u}}^{\star 0}_{i}
	= \vec{{\sigma}}^{\star0}_{i} \tilde{u}\bigg|_{{\sigma}^{\star0}_{i}}
, \text{ and}\\[2ex]
\flat^{\star1}\tilde{u}  = {\mathtt{u}}^{\star 1}_{i}
	=\left(\sum_{\forall \sigma^{\star0}_{j}\prec{\sigma}^{\star1}_{i}}\left|{\sigma}^{\star1}_{i} \cap \sigma^{3}_{j}\right|\, \tilde{u}\bigg|_{{\sigma}^{\star0}_{j}}\right)\cdot\vec{{\sigma}}^{\star1}_{i} 
	\approx \int_{{{\sigma}}^{\star1}_{i}} \tilde{u}\cdot d\vec{{\sigma}}^{\star1}_{i}.
\end{array}
\end{equation}
Note that musical isomorphisms are often typeset as $\tilde{u}^{\flat^{0}}$. The alternate typesetting $\flat^{0}\tilde{u}$ is chosen in this article to highlight order of operation and to simplify the presentation. In the definition above $\tilde{u}\big|_{{\sigma}^0}$ denotes the vector field evaluated at a given vertex ($0$-simplex), ${\sigma}^j\prec{\sigma}^k$ denotes all lower dimension $j$-simplices which are components of a given $k$-simplex, and $\left|{\sigma}^j \cap \sigma^{\star k}\right|$ is the portion of the volume of a $j$-simplex that is in a given $k$-cell. The directions $\vec{{\sigma}}^k$ are either sources ``$+1$'' or sinks ``$-1$'' for $0$-simplices, whereas for $1$-simplices it is the direction along the simplex. Typically, orientations are assigned assuming $0$-simplices are sources, but this is not necessary. 

The discrete flat operations $\flat^k$ and $\flat^{\star k}$ approximate the underlying vector field as constant within each dual or primal volume, respectively. Discrete flats of vector fields yield vector-valued $0$-cochains and scalar-valued $1$-cochains. The discrete flats to $1$-cochains increase the spatial dimension of the value by one. For example, if the underlying vector field has spatial dimension $[m]$, then the resulting $1$-cochain will have spatial dimension $[m^2]$. 

These results can be extended for tensor fields. For example, consider the tensor field  $\tilde{Q} \in \mathbb{R}^{3\times3}$ and the discrete flat to a vector-valued 1-cochain. The discrete flat is applied row-by-row to the tensor field to yield each component of the 1-cochain. Each row operation is written as:

\begin{equation}
	\flat^{1}\tilde{Q}_{j:} = {\mathtt{q}}^{1}_{i,j}
	\approx\int_{\vec{{\sigma}}_i^1} 
	\left[
	\begin{array}{ccc}
	\tilde{Q}_{j1} & \tilde{Q}_{j2} & \tilde{Q}_{j3}
	\end{array}\right]
	\cdot d\vec{{\sigma}}_i^1
	= \int_{\vec{{\sigma}}_i^1} \tilde{Q}_{j:}\cdot d\vec{{\sigma}}_i^1.
\end{equation}
Combining all the rows of the tensor, the resulting vector-valued 1-cochain is:

\begin{equation}
	\flat^{1}\tilde{Q}
	= {\mathtt{q}}_i^{1} =
	 \left[\begin{array}{c}
	{\mathtt{q}}^{1}_{i,1} \\[1ex]
	{\mathtt{q}}^{1}_{i,2} \\[1ex]
	{\mathtt{q}}^{1}_{i,3} 
	\end{array}\right] 
	\approx \left[\begin{array}{c}
	\int_{\vec{{\sigma}}_i^1} \tilde{Q}_{1:}\cdot d\vec{{\sigma}}_i^1 \\[1ex]
	\int_{\vec{{\sigma}}_i^1} \tilde{Q}_{2:}\cdot d\vec{{\sigma}}_i^1 \\[1ex]
	\int_{\vec{{\sigma}}_i^1} \tilde{Q}_{3:}\cdot d\vec{{\sigma}}_i^1 
	\end{array}\right] 
	= \int_{\vec{{\sigma}}_i^1} \tilde{Q}d\vec{{\sigma}}_i^1.
\end{equation}

\subsubsection{Discrete sharp musical isomorphism}

Discrete sharp ${{\sharp}}$ musical isomorphisms are maps from cochains back to vector fields. As such, they are inverse maps to discrete $\flat$ musical isomorphisms. As with the discrete flats, multiple discrete sharps have been proposed \cite{hirani:thesis}. It is also common to use Whitney maps \cite{whitney:1957}. The discrete sharps used in this article approximate the fields at the locations of the $0$-simplices or $0$-cells. 

For $0$-cochains, the discrete sharps used are exact inverse maps to the flats defined above, recovering the field exactly at the locations of the $0$-simplices or $0$-cells. As an example for vector-valued 0-cochains, they are defined as follows:
\begin{equation}
{{\sharp}}^{0}{\mathtt{u}}^{0}_{i} = {\mathtt{u}}^{0}_{i} = \tilde{u}\bigg|_{{\sigma}^{0}_{i}}
,\qquad \text{and} \qquad
{{\sharp}}^{\star0}{\mathtt{u}}^{\star 0}_{i} = {\mathtt{u}}^{\star 0}_{i} = \tilde{u}\bigg|_{{\sigma}^{\star0}_{i}}.
\end{equation}

The discrete sharps ${{\sharp}}$ for 1-cochains are, however, not exact inverses for the field evaluated at the locations of the $0$-simplices or $0$-cells. While, it may be possible in some circumstances to construct a left Moore-Penrose pseudo inverse of the discrete flats described earlier, the resulting discrete sharp ${{\sharp}}$ will require information from all $1$-simplices in the complex, in general. In other words, the operation is global. Furthermore, our experience is that this pseudo inverses are not well conditioned. Therefore, we construct a local left Moore-Penrose pseudo inverse for each element. This is an approximation, but keeps information local and is more well conditioned. For scalar-valued 1-cochains, the discrete sharps are:
\begin{equation}
\begin{array}{l}
{{\sharp}}^{1}{\mathtt{u}}^{1}_{j |\, \forall \sigma^{1}_{j}\prec{\sigma}^{0}_{i} }
	= (A^T\,A)^{-1} A^T\,{\mathtt{u}}^{1}_{j |\, \forall \sigma^{1}_{j}\prec{\sigma}^{0}_{i} }
	= \tilde{u}\bigg|_{{\sigma}^{0}_{i}};\quad 
A = \left[ \text{rows: } \vec{{\sigma}}^{1}_{j} \big|\, \forall \sigma^{1}_{j}\prec{\sigma}^{0}_{i} \right]
,\text{ and}\\[3ex]
{{\sharp}}^{\star1}{\mathtt{u}}^{\star 1}_{j |\, \forall \star \sigma^{1}_{j}\prec\star \sigma^{0}_{i} }
	= ({A}^T\,{A})^{-1} {A}^T\,{\mathtt{u}}^{\star 1}_{j |\, \forall \star \sigma^{1}_{j}\prec\star \sigma^{0}_{i} }
	= \tilde{u}\bigg|_{\star \sigma^{0}_{i}};\quad 
{A} = \left[ \text{rows: } \vec{{\sigma}}^{\star1}_{j} \big|\, \forall \star \sigma^{1}_{j}\prec\star \sigma^{0}_{i} \right]
.
\end{array}
\end{equation}
Here ${\sigma}^j\prec{\sigma}^k$ denotes all higher dimension $j$-simplices which are connected to a given $k$-simplex. Notice that the sharps for $1$-cochains lower the spatial order of the result by one, opposite to the discrete flats for $1$-cochains. 

The extension to vector valued $1$-cochains is straight forward by applying the discrete sharp to each component of the $1$-cochain independently. For the $k^{th}$ component
\begin{equation}
{{\sharp}}^{1}{\mathtt{q}}^{1}_{j |\, \forall \sigma^{1}_{j}\prec{\sigma}^{0}_{i},k}
	= (A^T\,A)^{-1} A^T\,{\mathtt{q}}^{1}_{j |\, \forall \sigma^{1}_{j}\prec{\sigma}^{0}_{i},k}
	= \left. \left[
	\begin{array}{ccc}
		\tilde{Q}_{k1}	&	\tilde{Q}_{k2}	&	\tilde{Q}_{k3} 
	\end{array}
	\right]\right|_{{\sigma}^{0}_{i}} 
	= \tilde{Q}_{k:}\bigg|_{{\sigma}^{0}_{i}}.
\end{equation}
Combining the three components of the vector valued primal $1$-cochain, we obtain the tensor field
\begin{equation}
{{\sharp}}^{1}{\mathtt{q}}^{1}_{j |\, \forall \sigma^{1}_{j}\prec{\sigma}^{0}_{i} }
  = \left[
	\begin{array}{c}
		{{\sharp}}^{1}{\mathtt{q}}^{1}_{j |\, \forall \sigma^{1}_{j}\prec{\sigma}^{0}_{i},1} \\[1ex]
		{{\sharp}}^{1}{\mathtt{q}}^{1}_{j |\, \forall \sigma^{1}_{j}\prec{\sigma}^{0}_{i},2} \\[1ex]
		{{\sharp}}^{1}{\mathtt{q}}^{1}_{j |\, \forall \sigma^{1}_{j}\prec{\sigma}^{0}_{i},3} 
	\end{array}
	\right] \approx \left. \left[
	\begin{array}{c}
		\tilde{Q}_{1:} \\[1ex]
		\tilde{Q}_{2:} \\[1ex]
		\tilde{Q}_{3:} 
	\end{array}
	\right]\right|_{{\sigma}^{0}_{i}}  = \tilde{Q}\bigg|_{{\sigma}^{0}_{i}} .
\end{equation}

\subsection{Connection to Vector Calculus}

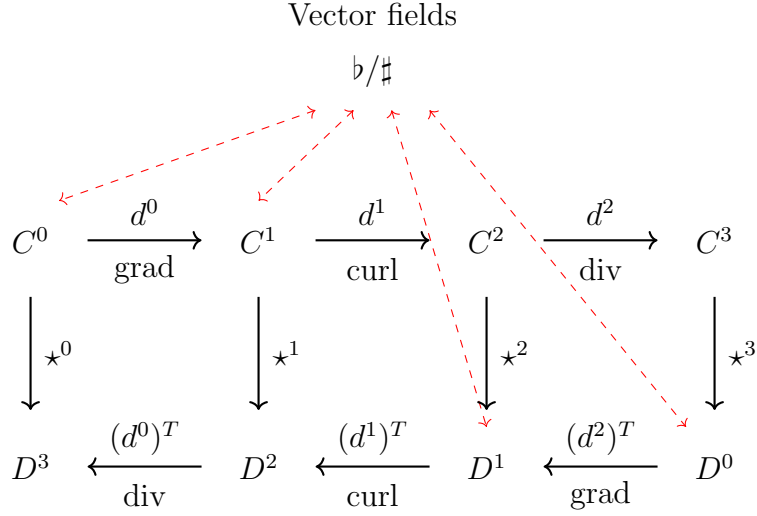
\begin{figure}[!t]
\centering
\begin{tikzpicture}[scale=1.5]
\draw (3,2) node {Vector fields};
\draw (3,1.5) node {$\flat/{{\sharp}}$};
\draw[dashed,<->,red] (2.5,1.15) -- (0.25,0.35);
\draw[dashed,<->,red] (2.833333,1.15) -- (2,0.35);
\draw[dashed,<->,red] (3.166666,1.15) -- (4,-1.65);
\draw[dashed,<->,red] (3.5,1.15) -- (5.75,-1.65);

\draw (0,0) node {$C^0$};
\draw (1,0.25) node {$d^0$};
\draw (1,-0.25) node {grad};
\draw[thick, ->] (0.5,0) -- (1.5,0);
\draw (2,0) node {$C^1$};
\draw (3,0.25) node {$d^1$};
\draw (3,-0.25) node {curl};
\draw[thick, ->] (2.5,0) -- (3.5,0);
\draw (4,0) node {$C^2$};
\draw (5,0.25) node {$d^2$};
\draw (5,-0.25) node {div};
\draw[thick, ->] (4.5,0) -- (5.5,0);
\draw (6,0) node {$C^3$};

\draw[thick, ->] (0,-0.5) -- (0,-1.5);
\draw (0.25,-1) node {$\star^0$};
\draw (0,-2) node {$D^3$};
\draw (1,-2+0.25) node {$(d^{0})^T$};
\draw (1,-2-0.25) node {div};
\draw[thick, <-] (0.5,-2) -- (1.5,-2);
\draw[thick, ->] (2,-0.5) -- (2,-1.5);
\draw (2+0.25,-1) node {$\star^1$};
\draw (2,-2) node {$D^2$};
\draw (3,-2+0.25) node {$(d^{1})^T$};
\draw (3,-2-0.25) node {curl};
\draw[thick, <-] (2.5,-2) -- (3.5,-2);
\draw[thick, ->] (4,-0.5) -- (4,-1.5);
\draw (4+0.25,-1) node {$\star^2$};
\draw (4,-2) node {$D^1$};
\draw (5,-2+0.25) node {$(d^{2})^T$};
\draw (5,-2-0.25) node {grad};
\draw[thick, <-] (4.5,-2) -- (5.5,-2);
\draw[thick, ->] (6,-0.5) -- (6,-1.5);
\draw (6+0.25,-1) node {$\star^3$};
\draw (6,-2) node {$D^0$};
\end{tikzpicture}
\caption{Relationship between de Rham complex, vector fields, and vector calculus operations. \label{fig:vderham}}
\end{figure}

With the knowledge that DEC satisfies Stokes theorem, and using the discrete musical isomorphisms, it is possible to associate different discrete exterior derivatives with the traditional vector calculus operations: gradient, curl, and divergence. These relationships mirror those in smooth exterior calculus and are summarised mathematically below:
\begin{equation}
\begin{array}{llll}
\text{grad }\tilde{u} &= \nabla \tilde{u} 
	&\approx {{\sharp}}^{1}d^0 \flat^{0}\tilde{u} 
	&= {{\sharp}}^{1}d^0 {\mathtt{u}}^{0} \\
	&& \approx {{\sharp}}^{\star1}(d^2)^T \flat^{\star0}\tilde{u} 
	&= {{\sharp}}^{\star1}(d^2)^T {\mathtt{u}}^{\star 0} \\
\text{curl }\tilde{u}  &= \nabla\times \tilde{u} 
	&\approx {{\sharp}}^{\star1}\star^2 d^1 \flat^{1}\tilde{u} 
	&= {{\sharp}}^{\star1}\star^2 d^1 {\mathtt{u}}^{1} \\
	&& \approx {{\sharp}}^{1}(\star^1)^{-1} (d^1)^T \flat^{\star1}\tilde{u}
	&= {{\sharp}}^{1}(\star^1)^{-1} (d^1)^T {\mathtt{u}}^{\star 1} \\
\text{div }\tilde{u} &= \nabla\cdot \tilde{u} 
	&\approx {{\sharp}}^{0}\star^{-1}_0 (d^0)^T \star^1 \flat^{1}\tilde{u}
	&= {{\sharp}}^{0}\star^{-1}_0 (d^0)^T \star^1 {\mathtt{u}}^{1} \\
	&& \approx {{\sharp}}^{\star0}\star^3 d^2 (\star^2)^{-1} \flat^{\star1}\tilde{u}
	&= {{\sharp}}^{\star0}\star^3 d^2 (\star^2)^{-1} {\mathtt{u}}^{\star 1}
\end{array}
\end{equation}
This is also shown pictorially in Figure \ref{fig:vderham}.

%
\section{Linear elasticity using discrete exterior calculus}\label{sec:linDEC}
%

This section applies discrete exterior calculus to the equations of linear elasticity. The first task is to approximate the solid body of interest with a primal simplicial complex, along with a Voronoi dual. The primary values of interest in solid mechanics are deformations of the body and the resulting stresses. Under the infinitesimal strain assumption, it is common to use the displacement field $\tilde{u}$ to described the deformation of the body. Given that the geometries of the primal and dual complexes are defined entirely by the vertices of the primal complex and their connections, it is natural to map the displacements there: ${\mathtt{u}}^{0} = \flat^{0}\tilde{u}$ . Thus the displacement field becomes a vector-valued primal $0$-cochain, similar to the approach of Yavari \cite{yavari:2008}.

In order to apply the constitutive law, the infinitesimal strain tensor must be evaluated, which is a function of the displacement gradient. The traditional vector calculus gradient is computed using discrete exterior calculus as:
\begin{equation}
	\nabla \tilde{u} = {{\sharp}}^{1}d^0 \flat^{0}\tilde{u} = {{\sharp}}^{1}d^0 {\mathtt{u}}^{0}.
\end{equation}
Notice that the gradient is computed from the displacement $0$-cochain. Now the infinitesimal strain tensor can be constructed 
\begin{equation}
	\tilde{\epsilon} = \frac{1}{2}\left( {{\sharp}}^{1}d^0 {\mathtt{u}}^{0} + \left( {{\sharp}}^{1}d^0 {\mathtt{u}}^{0}\right)^T\right)
	 = \frac{1}{2} {P} {{\sharp}}^{1}d^0 {\mathtt{u}}^{0},
\end{equation} 
where, assuming the tensor is arranged as vector $\tilde{\epsilon} = [\tilde{\epsilon}_{xx}\ \tilde{\epsilon}_{xy}\ \tilde{\epsilon}_{xz}\ \tilde{\epsilon}_{yx}\ \tilde{\epsilon}_{yy}\ \tilde{\epsilon}_{yz}\ \tilde{\epsilon}_{zx}\ \tilde{\epsilon}_{zy}\ \tilde{\epsilon}_{zz}\ ]^T$,  $P_\text{grad}$ is permutation matrix:
\begin{equation}
	P_\text{grad} = 
	\begin{bsmallmatrix}
		2 & 0 & 0 & 0 & 0 & 0 & 0 & 0 & 0 \\
		0 & 1 & 0 & 1 & 0 & 0 & 0 & 0 & 0 \\
		0 & 0 & 1 & 0 & 0 & 0 & 1 & 0 & 0 \\
		0 & 1 & 0 & 1 & 0 & 0 & 0 & 0 & 0 \\
		0 & 0 & 0 & 0 & 2 & 0 & 0 & 0 & 0 \\
		0 & 0 & 0 & 0 & 0 & 1 & 0 & 1 & 0 \\
		0 & 0 & 1 & 0 & 0 & 0 & 1 & 0 & 0 \\
		0 & 0 & 0 & 0 & 0 & 1 & 0 & 1 & 0 \\
		0 & 0 & 0 & 0 & 0 & 0 & 0 & 0 & 2 
	\end{bsmallmatrix}.
\end{equation}
Alternatively, this is the point where the deformation gradient could be constructed to apply finite strain theories. However, in this article we restrict our interest to the linear theory. 

Finally, the constitutive law can be applied in a similar fashion using the discrete displacement gradient directly and another permutation matrix:
\begin{equation}
	\tilde{\tau} = \lambda \text{tr}(\tilde{\epsilon}) + 2\mu\tilde{\epsilon} =  P_{\tilde{\tau}} {{\sharp}}^{1}d^0 {\mathtt{u}}^{0},
\end{equation}
where $\text{tr}$ is the trace, and the permutation matrix $P_{\tilde{\tau}}$ is:
\begin{equation}
	P_{\tilde{\tau}} = 
	\begin{bsmallmatrix}
		2\mu + \lambda 	& 0 		& 0 		& 0 		& \lambda	& 0 		& 0 		& 0 		& \lambda \\
		0 			& \mu 	& 0 		& \mu 	& 0 			& 0 		& 0 		& 0 		& 0 \\
		0 			& 0 		& \mu 	& 0 		& 0 			& 0 		& \mu 	& 0 		& 0 \\
		0 			& \mu 	& 0 		& \mu 	& 0 			& 0 		& 0 		& 0 		& 0 \\
		\lambda	& 0 		& 0 		& 0 		& 2\mu + \lambda 	& 0 		& 0 		& 0 		& \lambda \\
		0 			& 0 		& 0 		& 0 		& 0 			& \mu 	& 0 		& \mu 	& 0 \\
		0 			& 0 		& \mu 	& 0 		& 0 			& 0 		& \mu 	& 0 		& 0 \\
		0 			& 0 		& 0 		& 0 		& 0 			& \mu 	& 0 		& \mu 	& 0 \\
		\lambda	& 0 		& 0 		& 0 		& \lambda	& 0 		& 0 		& 0 		& 2\mu + \lambda 
	\end{bsmallmatrix}.
\end{equation}

The last step is to apply the balance of linear momentum. Recall that the constitutive law is constructed such that the balance of angular momentum is satisfied in the continuum. The static balance of linear momentum is evaluated by taking the divergence of the stress field:
\begin{equation}
	\nabla\cdot \tilde{\tau} +\tilde{f} = {{\sharp}}^{0}(\star^0)^{-1} (d^0)^T \star^1 \flat^{1}P_{\tilde{\tau}} {{\sharp}}^{1}d^0 \flat^{0}\tilde{u} +\tilde{f}=0.
\end{equation}
This can be evaluated using discrete exterior calculus without reference to traditional vector calculus by removing the fist and last musical isomorphisms:
\begin{equation}
	(\star^0)^{-1} (d^0)^T \star^1 \flat^{1}P_{\tilde{\tau}} {{\sharp}}^{1}d^0 {\mathtt{u}}^0 + \mathtt{f}_0=0, \label{eq:DEC_BLM}
\end{equation}
where $\mathtt{f}_0 = \flat^0\tilde{f}$. The DEC formalism considers the Hodge stars as incorporating both geometric and physical properties, which can be accomplished as follows:
\begin{equation}
	\star^{1}_{\text{mat}} =  \star^1 \flat^{1}P_{\tilde{\tau}} {{\sharp}}^{1},
\end{equation}
and the resulting balance of linear momentum becomes:
\begin{equation}
	(\star^0)^{-1} (d^0)^T \star^{1}_{\text{mat}} d^0 {\mathtt{u}}^{0}  + \mathtt{f}_0 = 0.
\end{equation}
This is the Laplace equation with non-local and non-diagonal material Hodge star.

\subsection{Simplified formulation}

Observe that one component of the permutation matrix $P$ is the shear modulus multiplied by the identity matrix:
\begin{equation} 
	P_{\tilde{\tau}} = \mu I + {P}_{\tilde{\tau}^\prime} = \mu I + 
	\begin{bsmallmatrix}
		\mu + \lambda 	& 0 		& 0 		& 0 		& \lambda	& 0 		& 0 		& 0 		& \lambda \\
		0 			& 0 	& 0 		& \mu 	& 0 			& 0 		& 0 		& 0 		& 0 \\
		0 			& 0 		& 0 	& 0 		& 0 			& 0 		& \mu 	& 0 		& 0 \\
		0 			& \mu 	& 0 		& 0 	& 0 			& 0 		& 0 		& 0 		& 0 \\
		\lambda	& 0 		& 0 		& 0 		& \mu + \lambda 	& 0 		& 0 		& 0 		& \lambda \\
		0 			& 0 		& 0 		& 0 		& 0 			& 0 	& 0 		& \mu 	& 0 \\
		0 			& 0 		& \mu 	& 0 		& 0 			& 0 		& 0 	& 0 		& 0 \\
		0 			& 0 		& 0 		& 0 		& 0 			& \mu 	& 0 		& 0 	& 0 \\
		\lambda	& 0 		& 0 		& 0 		& \lambda	& 0 		& 0 		& 0 		& \mu + \lambda 
	\end{bsmallmatrix}.
\end{equation}
Furthermore, recall that ideally the sharp and flat musical isomorphisms are exact inverses. Therefore, the material Hodge star can be rewritten as:
\begin{equation} 
	\star^{1}_{\text{mat}} =  \star^1 (\mu + \flat^{1}{P}_{\tilde{\tau}^\prime} {{\sharp}}^{1}),
\end{equation}
which minimises the use of approximate musical isomorphisms. Another possibility, is to partition the constitutive law into shape change, volume change, and rotations:
\begin{align}
	P_{\tilde{\tau}} &= (2\mu I)_\text{shape} + P_\text{vol} + P_\text{rot}\\&= 2\mu I + 
	\begin{bsmallmatrix}
		\lambda 	& 0 		& 0 		& 0 		& \lambda	& 0 		& 0 		& 0 		& \lambda \\
		0 				& 0 		& 0 		& 0 		& 0 			& 0 		& 0 		& 0 		& 0 \\
		0 				& 0 		& 0 		& 0 		& 0 			& 0 		& 0 		& 0 		& 0 \\
		0 				& 0 		& 0 		& 0 		& 0 			& 0 		& 0 		& 0 		& 0 \\
		\lambda	& 0 		& 0 		& 0 		& \lambda 	& 0 		& 0 		& 0 		& \lambda \\
		0 				& 0 		& 0 		& 0 		& 0 				& 0 		& 0 		& 0 		& 0 \\
		0 				& 0 		& 0 		& 0 		& 0 				& 0 		& 0 		& 0 		& 0 \\
		0 				& 0 		& 0 		& 0 		& 0 				& 0 		& 0 		& 0 		& 0 \\
		\lambda	& 0 		& 0 		& 0 		& \lambda	& 0 		& 0 		& 0 		& \lambda 
	\end{bsmallmatrix} +
	\begin{bsmallmatrix}
		0 		& 0 		& 0 		& 0 		& 0		& 0 		& 0 		& 0 		& 0 \\
		0 		& -\mu 	& 0 		& \mu 	& 0 		& 0 		& 0 		& 0 		& 0 \\
		0 		& 0 		& -\mu 	& 0 		& 0 		& 0 		& \mu 	& 0 		& 0 \\
		0 		& \mu 	& 0 		& -\mu 	& 0 		& 0 		& 0 		& 0 		& 0 \\
		0		& 0 		& 0 		& 0 		& 0 		& 0 		& 0 		& 0 		& 0 \\
		0 		& 0 		& 0 		& 0 		& 0 		& -\mu 	& 0 		& \mu 	& 0 \\
		0 		& 0 		& \mu 	& 0 		& 0 		& 0 		& -\mu 	& 0 		& 0 \\
		0 		& 0 		& 0 		& 0 		& 0 		& \mu 	& 0 		& -\mu	& 0 \\
		0		& 0 		& 0 		& 0 		& 0		& 0 		& 0 		& 0 		& 0
	\end{bsmallmatrix},
\end{align}
which is formulation is adopted in the simulation presented below.

\subsection{Closure for discrete divergence and Neumann boundary conditions}

The discrete exterior derivative $(d^0)^T$ used in the divergence of stresses sums values from dual $2$-cochains (dual faces) to dual $3$-cochains (dual volumes). However, the dual $3$-cells at the surface of the geometry are truncated by outer bounds of the primal complex, and are therefore incomplete. In other words, the dual $3$-cells that intersect with the surface of the mesh are not closed by dual $2$-cells. In order to have an accurate value for the divergence in these $3$-cells, an additional closure is required.

The closure used in this article is inspired by the closure presented in Mohamed \textit{et al} \cite{mohamed:2016}. It is constructed on sub-patches of the surface primal $2$-simplices (primal faces) where the dual $3$-cells are truncated. The decomposition of surface primal $2$-simplices is shown in Figure \ref{fig:subfaces}. The closure is built from a discrete surface flat $\flat^{{1s}}$ and discrete surface Hodge star $\star^{1s}$ of the stress field evaluated at primal $0$-simplices to the sub-patches of the surface primal $2$-simplices:
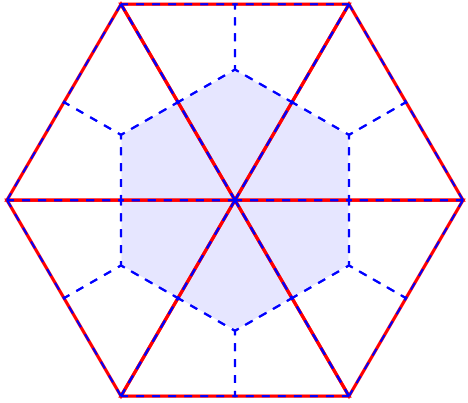
\begin{figure}[!t]
\centering
\begin{tikzpicture}[scale=3]

\filldraw[fill = blue!10!white, draw=white]  (0.5,1/3*0.86602540378) -- (0.5,1/3*-0.86602540378) -- (1,2/3*-0.86602540378) -- (1.5,1/3*-0.86602540378) -- (1.5,1/3*0.86602540378) -- (1,2/3*0.86602540378) -- cycle;

\draw[very thick, red] (0,0) -- (0.5,0.86602540378) -- (1,0) --cycle;
\draw[very thick, red] (0,0) -- (0.5,-0.86602540378) -- (1,0) --cycle;
\draw[very thick, red] (1,0) -- (0.5,0.86602540378) -- (1.5,0.86602540378) --cycle;
\draw[very thick, red] (1,0) -- (0.5,-0.86602540378) -- (1.5,-0.86602540378) --cycle;
\draw[very thick, red] (2,0) -- (1.5,0.86602540378) -- (1,0) --cycle;
\draw[very thick, red] (2,0) -- (1.5,-0.86602540378) -- (1,0) --cycle;

\draw[draw=blue,dashed,thick]  (0,0) -- (0.25,0.5*0.86602540378) -- (0.5,1/3*0.86602540378)  -- (0.5,0) --cycle;
\draw[draw=blue,dashed,thick]  (0.5,0.86602540378) -- (0.25,0.5*0.86602540378) -- (0.5,1/3*0.86602540378)  -- (0.75,0.5*0.86602540378) --cycle;
\draw[draw=blue,dashed,thick]  (1,0)  -- (0.75,0.5*0.86602540378) -- (0.5,1/3*0.86602540378) -- (0.5,0) --cycle;

\draw[draw=blue,dashed,thick]  (1,0) -- (1.25,0.5*0.86602540378) -- (1.5,1/3*0.86602540378)  -- (1.5,0) --cycle;
\draw[draw=blue,dashed,thick]  (1.5,0.86602540378) -- (1.25,0.5*0.86602540378) -- (1.5,1/3*0.86602540378)  -- (1.75,0.5*0.86602540378) --cycle;
\draw[draw=blue,dashed,thick]  (2,0)  -- (1.75,0.5*0.86602540378) -- (1.5,1/3*0.86602540378) -- (1.5,0) --cycle;

\draw[draw=blue,dashed,thick]  (1,0) -- (0.75,0.5*0.86602540378) -- (1,2/3*0.86602540378)  -- (1.25,0.5*0.86602540378) --cycle;
\draw[draw=blue,dashed,thick]  (0.5,0.86602540378) -- (0.75,0.5*0.86602540378) -- (1,2/3*0.86602540378)  -- (1,0.86602540378) --cycle;
\draw[draw=blue,dashed,thick]  (1.5,0.86602540378) -- (1.25,0.5*0.86602540378) -- (1,2/3*0.86602540378)  -- (1,0.86602540378) --cycle;

\draw[draw=blue,dashed,thick]  (1,0) -- (0.75,0.5*-0.86602540378) -- (1,2/3*-0.86602540378)  -- (1.25,0.5*-0.86602540378) --cycle;
\draw[draw=blue,dashed,thick]  (0.5,-0.86602540378) -- (0.75,0.5*-0.86602540378) -- (1,2/3*-0.86602540378)  -- (1,-0.86602540378) --cycle;
\draw[draw=blue,dashed,thick]  (1.5,-0.86602540378) -- (1.25,0.5*-0.86602540378) -- (1,2/3*-0.86602540378)  -- (1,-0.86602540378) --cycle;

\draw[draw=blue,dashed,thick]  (0,0) -- (0.25,0.5*-0.86602540378) -- (0.5,1/3*-0.86602540378)  -- (0.5,0) --cycle;
\draw[draw=blue,dashed,thick]  (0.5,-0.86602540378) -- (0.25,0.5*-0.86602540378) -- (0.5,1/3*-0.86602540378)  -- (0.75,0.5*-0.86602540378) --cycle;
\draw[draw=blue,dashed,thick]  (1,0)  -- (0.75,0.5*-0.86602540378) -- (0.5,1/3*-0.86602540378) -- (0.5,0) --cycle;

\draw[draw=blue,dashed,thick]  (1,0) -- (1.25,0.5*-0.86602540378) -- (1.5,1/3*-0.86602540378)  -- (1.5,0) --cycle;
\draw[draw=blue,dashed,thick]  (1.5,-0.86602540378) -- (1.25,0.5*-0.86602540378) -- (1.5,1/3*-0.86602540378)  -- (1.75,0.5*-0.86602540378) --cycle;
\draw[draw=blue,dashed,thick]  (2,0)  -- (1.75,0.5*-0.86602540378) -- (1.5,1/3*-0.86602540378) -- (1.5,0) --cycle;
\end{tikzpicture}
\caption{Closure for the divergence discrete exterior derivative $(d^0)^T$. The figure shows a surface patch of six primal $2$-simplices outlined in red. The sub-patch surface primal $2$-simplices associated with the central $0$-simplex is outlined with dashed blue lines and filled with a lighter blue.  \label{fig:subfaces}}
\end{figure}
\begin{equation}
	\star^{1s}_{i,j} = \frac{|{\sigma}^{2}_{i,j}|}{|{\sigma}^{\star1}_{j}|}, \quad\text{ and }\quad
	\flat^{1s}_{i,j}\tilde{\tau} = \tilde{\tau}\bigg|_{{\sigma}^{0}_{i}}
	\cdot\vec{{\sigma}}^{\star1}_{j}, \quad \forall \sigma^{2}_{j}\prec{\sigma}^{0}_{i},
\end{equation}
where the superscript $s$ denotes operators for the surface closure of the dual $3$-cells, and ${\sigma}^{2}_{j}\prec{\sigma}^{0}_{i}$ denotes all surface $2$-simplices which are connected to a given $0$-simplex. Recall $3$-cells are dual to $0$-simplices in the present case. Each sub-patch of a surface primal $2$-simplex has area $|{\sigma}^{2}_{i,j}|$ and a dual $1$-cell direction $\vec{{\sigma}}^{\star1}_{j}$.

Finally, the discrete exterior derivative $(d^0)^T$ is closed by adding the additional contributions from the sub-patches of surface primal $2$-simplex, $(d^{0s})^T$,  closing each dual $3$-cell:
\begin{align}
	\nabla \tilde{\tau} &=  (\underbrace{{{\sharp}}^0 (\star^0)^{-1} (d^0)^T \star^1 \flat^1}_\text{standard divergence} + \underbrace{{{\sharp}}^0 (\star^0)^{-1} (d^{0s})^T \star^{1s} \flat^{1s} }_\text{closure}) \tilde{\tau}\\ 
	&= {{\sharp}}^0 (\star^0)^{-1} ((d^0)^T \star^1 \flat^1 + (d^{0s})^T \star^{1s} \flat^{1s} ) \tilde{\tau}.
\end{align}
Substituting the DEC definition of the stress tensor and simplifying, one obtains:
\begin{align}
	\nabla \tilde{\tau} &= {{\sharp}}^0 (\star^0)^{-1} \left[(d^0)^T \star^1 (2\mu + \flat^1 ( {P}_\text{vol}+{P}_\text{rot}) {{\sharp}}^1) + (d^{0s})^T \star^{1s} \flat^{1s}  P_{\tilde{\tau}} {{\sharp}}^1\right] d^0 {\mathtt{u}}^{0} \\
	 &= {{\sharp}}^0 (\star^0)^{-1} ((d^0)^T \star^{1}_{\text{mat}} +\, (d^{0s})^T \star^{1s}_{\text{mat}} )\, d^0 {\mathtt{u}}^{0} 
\end{align}
where $\star^{1s}_{\text{mat}} = \star^{1s} \flat^{1s} P_{\tilde{\tau}} {{\sharp}}^1$. 

The same discrete surface flat, discrete surface Hodge star, and divergence closure are used to enforce natural (Neuman) boundary conditions:
\begin{equation}
	\tilde{\tau}\cdot{n}\bigg|_{{\sigma}^{0}} = \tilde{t}\,\bigg|_{{\sigma}^{0}} = {{\sharp}}^0 (d^{0s})^T \star^{1s}_{\text{mat}} d^0 {\mathtt{u}}^{0}  = {{\sharp}}^0\mathtt{t}_{0s}.
\end{equation}

%
\section{Numerical results}\label{sec:numeric}
%

This section presents numerical results obtained using the theory described above. The goal is to demonstrate the application of the theory to a variety of common mechanical problems. The primal complexes are generated using TetGen \cite{tetgen} and the simulation software is implemented in Python 2.7, making use of the PyDEC library \cite{pydec}. 

The governing equations are solved in the form:
\begin{equation}
	\left[
	\begin{array}{cc}
		(\star^0)^{-1} (d^0)^T \star^{1}_{\text{mat}} d^0  & I_{0,{0s}}\\[1ex]
		(d^{0s})^T \star^{1s}_{\text{mat}} d^0 & -I_{n_{0s}}
	\end{array}
	\right]
	\left[
	\begin{array}{c}
		{\mathtt{u}}^0\\[1ex]
		\mathtt{t}_{0s}
	\end{array}
	\right]
	=
	\left[
	\begin{array}{c}
		\mathtt{0}_0\\[1ex]
		\mathtt{0}_{0s}
	\end{array}
	\right], \label{eq:sys}
\end{equation}
where $I_{0s}$ is an identity matrix with size equal to the number of surface $0$-simplices, and $I_{0,0s}$ is an identity matrix with the number of rows equal to the total number of $0$-simplices, and only the columns associated with surface $0$-simplices.

Boundary conditions are enforced by adding a source term to the left-hand side of equation \eqref{eq:sys} equal to the negative of the boundary condition times the appropriate column of the system matrix. The boundary values are removed from the solution vector, as well as the corresponding columns of the system matrix, and the final system is solved as an overdetermined problem.

\subsection{Compression of a circular cylinder}

\begin{figure}[!t]
\centering
\includegraphics[width=0.35\textwidth]{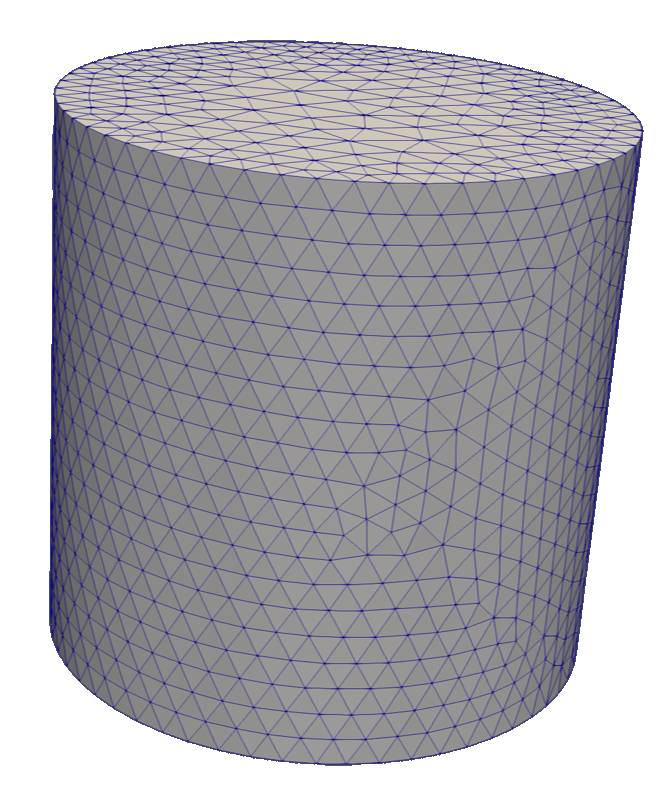}
\caption{Primal complex of circular cylinder \label{fig:cyl}}
\end{figure}

The first simulation is the compression of a circular cylinder in order to test the ability of the proposed theory to recover the full range Poisson's ratios. The test case is relatively simple, producing a constant uniaxial stress field with no shear components. Furthermore, none of the cells should rotate significantly in the deformed configuration.  The chosen cylinder geometry has a radius of $0.5$ and height of $1$. The primal complex shown in Figure \ref{fig:cyl} has $3,408$ vertices and $21,142$ edges. The compression is imposed by prescribing a displacement of $-0.05$ on the top surface. The Young's modulus is set to $209$ GPa and a range of Poisson's ratios from $-0.95$ to $0.45$ in $0.05$ increments are tested.  

\begin{figure}[!t]
\centering
\includegraphics[width=0.5\textwidth]{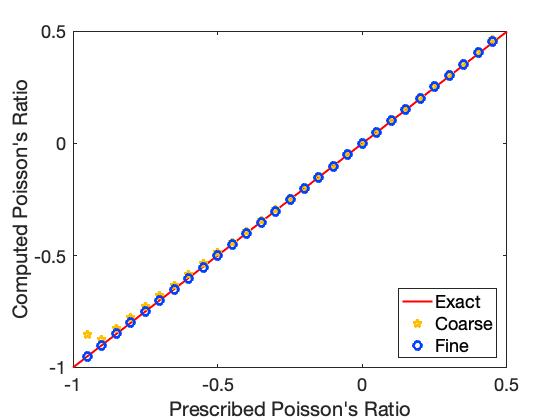}
\caption{Computed Poisson's ratio for the compression of a circular cylinder \label{fig:cyl_nu}}
\end{figure}

The computed Poisson's ratios are shown graphically in Figure \ref{fig:cyl_nu}. These values are computed as an average over all $0$-simplices at the outer radius of the cylinder. Figure \ref{fig:cyl_nu} also shows results from a coarse simplicial complex with $124$ vertices and $683$ edges. Apart from the extreme negative Poisson's ratios, where the results on the coarse grid are over predicted, the results are in good agreement with theory.  The components of the stress tensor for Poisson's ratio $0.3$ evaluated at every primal $0$-simplex are found in Table \ref{tab:ccstress}. These also show good agreement with theory, and are indicative of the accuracy obtained at other Poisson's ratios.

\begin{table}[!t]
\centering
\begin{tabular}{|c|c|c|}
\hline
 			& Mean value (GPa) 	& Standard Deviation \\
\hline
$\tilde{\tau}_{xx}$		& 0.0000			& 0.0005 \\
\hline
$\tilde{\tau}_{yy}$		& 0.0000			& 0.0005 \\
\hline
$\tilde{\tau}_{zz}$		& 10.4499 		& 0.0025 \\[-2ex]
    		& (exact: 10.45)		& \\
\hline
$\tilde{\tau}_{xy}$		& 0.0000			& 0.0001 \\
\hline
$\tilde{\tau}_{xz}$		& 0.0000			& 0.0009 \\
\hline
$\tilde{\tau}_{yz}$		& 0.0000			& 0.0006 \\
\hline
\end{tabular}
\caption{Components of the stress tensor computed for the compression of a cicular cylinder at Poisson's ratio $0.3$. \label{tab:ccstress}}
\end{table}

\subsection{Twist of a circular cylinder}
Now consider the twist of a circular cylinder, achieved by applying a prescribed rotation of $0.05$ radians to the top surface of the cylinder. Note that the shape of the upper and lower surfaces are fixed. In this case, the stress field still only has a single component: axial shear which varies in the radial direction. The simulations make use of the same geometry and complexes as the compression simulations. The Young's modulus and the sequence of Poisson's ratios tested are also the same.

\begin{figure}[!t]
\centering
\includegraphics[width=0.32\textwidth]{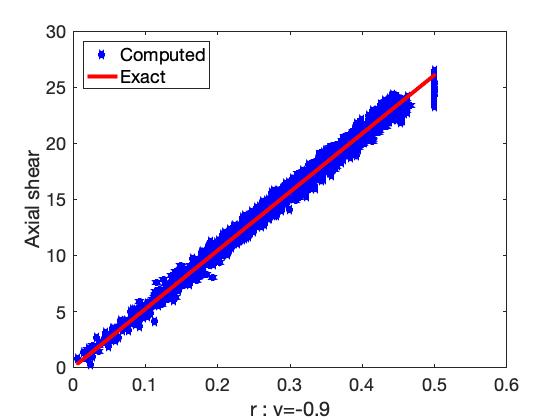}
\includegraphics[width=0.32\textwidth]{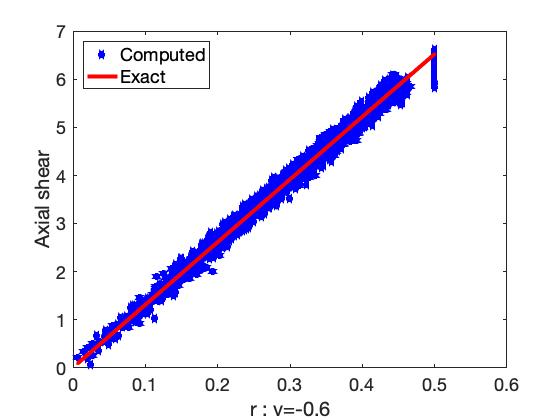}
\includegraphics[width=0.32\textwidth]{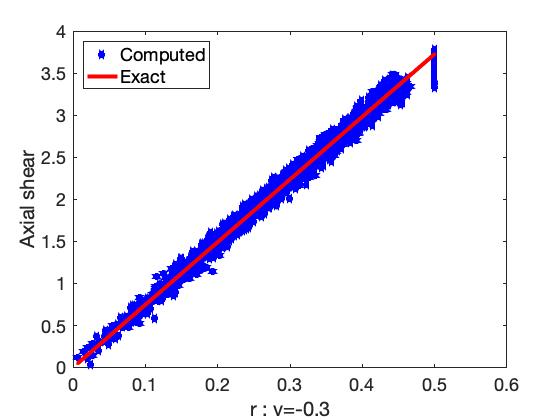}
\includegraphics[width=0.32\textwidth]{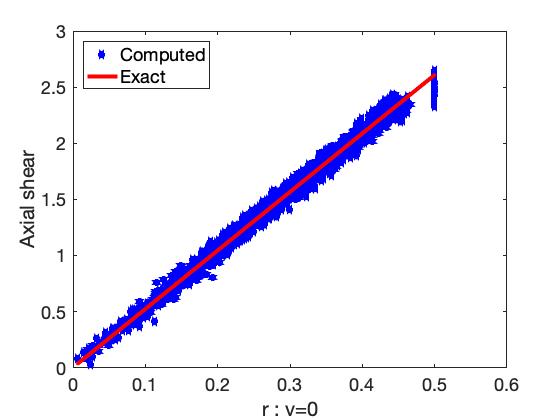}
\includegraphics[width=0.32\textwidth]{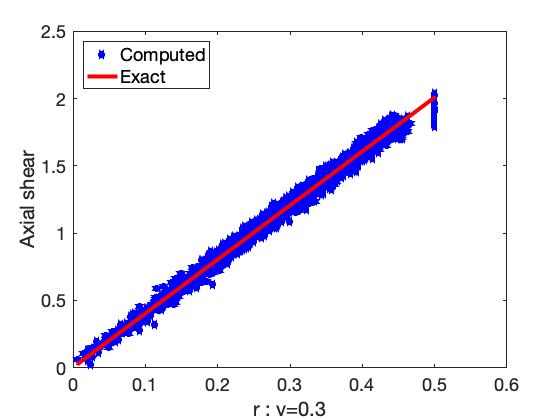}
\caption{Computed axial shear stress as a function of radius for the twist of a circular cylinder \label{fig:cyl_shear}}
\end{figure}

The computed axial shear stress as a function of radius is shown in Figure \ref{fig:cyl_shear}. While the computed values do not lie precisely on the exact solution, they are scatter about it. A possible reason for this error is the non-local nature of the discrete flat and sharp musical isomorphisms used to compute the stress tensor. Tracking the error in shear stress relative to the maximum exact value, the error drops from a maximum of $\sim23\%$ on the coarse mesh to a maximum of $\sim11\%$ on the finer mesh, where the cells are smaller and the computation becomes more local. The average errors also decrease from $\sim9.0\%$ and $\sim2.4\%$,  on the coarse and fine meshes, respectively.

\begin{figure}[!t]
\centering
\includegraphics[width=0.32\textwidth]{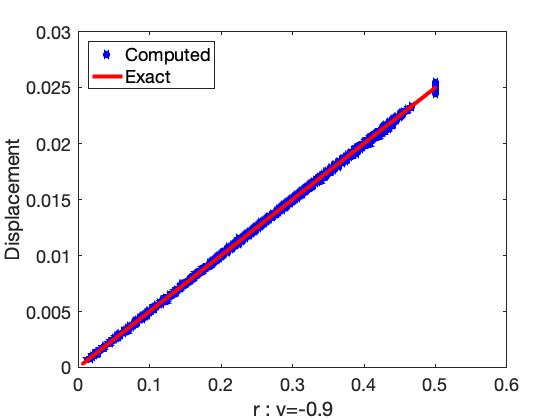}
\includegraphics[width=0.32\textwidth]{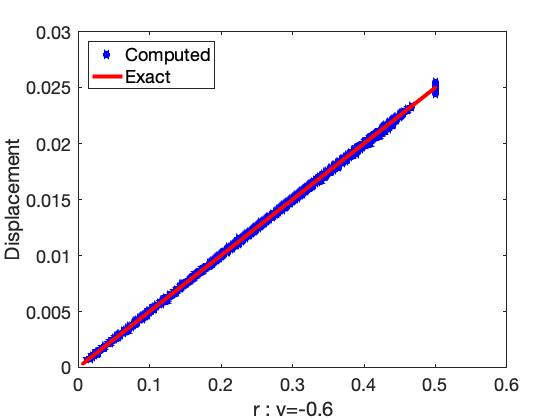}
\includegraphics[width=0.32\textwidth]{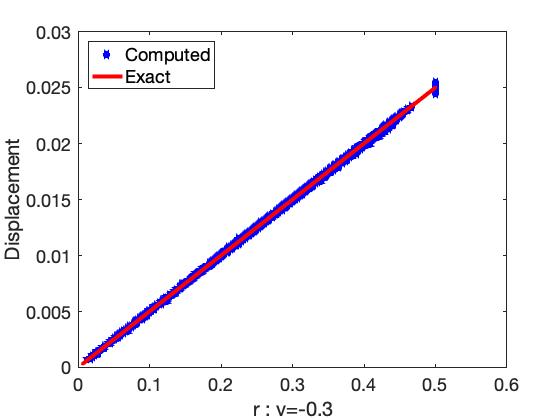}
\includegraphics[width=0.32\textwidth]{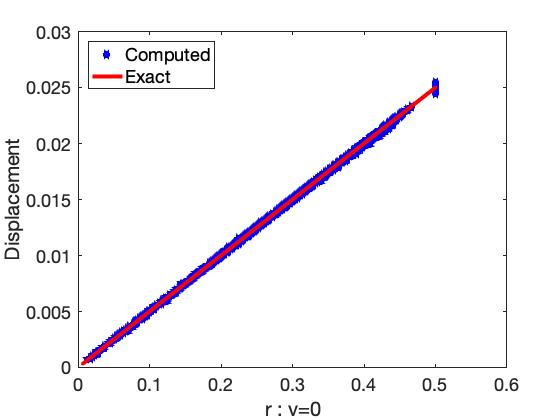}
\includegraphics[width=0.32\textwidth]{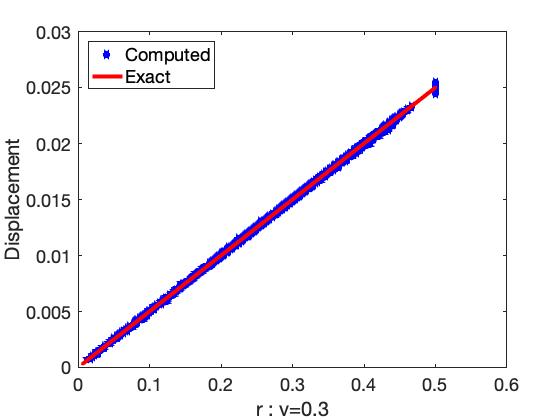}
\caption{Computed displacement magnitude divided by height as a function of radius for the twist of a circular cylinder \label{fig:cyl_disp}}
\end{figure}

In contrast to the stress field, the displacements are much more accurate as shown in Figure \ref{fig:cyl_disp}. The max error in displacements on the coarse grid is on the order of $\sim\%6$ and $\sim\%2$ on the fine mesh, with averages being about ten times lower.

\subsection{Point load on a cantilever beam}
\begin{figure}[!t]
\centering
\includegraphics[width=0.65\textwidth]{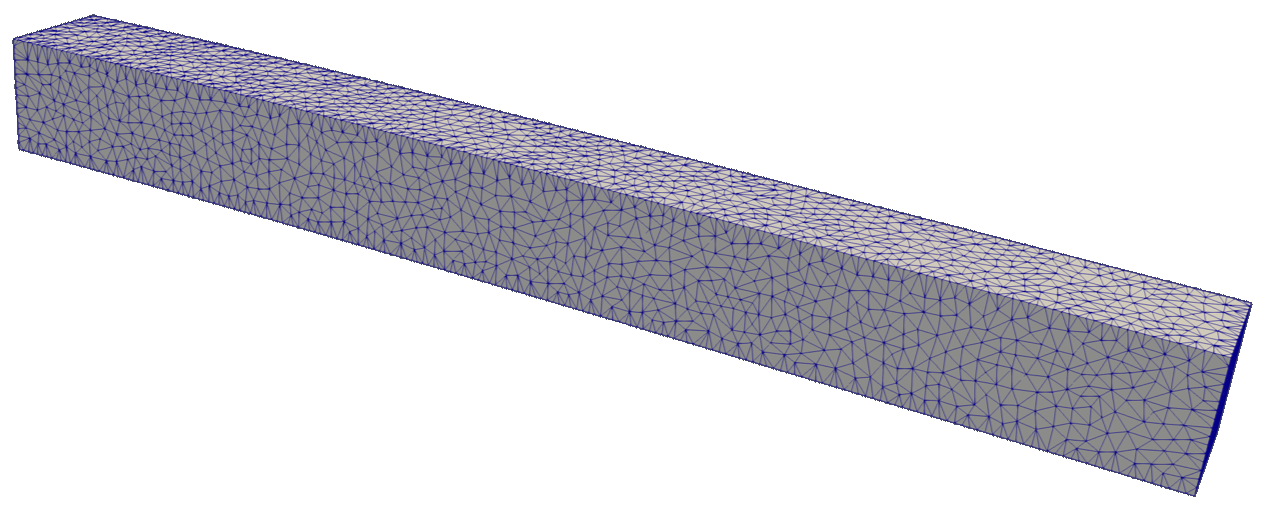}
\caption{Primal complex of cantilever beam \label{fig:beam}}
\end{figure}

Next, the deflection of a clamped square section cantilever beam with a single point load at the tip is simulated. In this case, the geometry has both a reasonable amount of displacement and rotation. The beam has a length of $10$ as well as both a height and depth of $1$. A displacement of $-0.19138756$ is prescribed to the upper surface at the tip of the beam. The geometry is approximated by a simplicial complex, shown in Figure \ref{fig:beam}, with $2,575$ nodes and $13,998$ edges.

\begin{figure}[!t]
\centering
\includegraphics[width=0.32\textwidth]{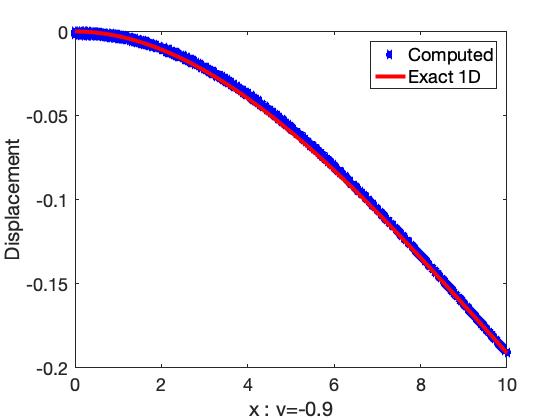}
\includegraphics[width=0.32\textwidth]{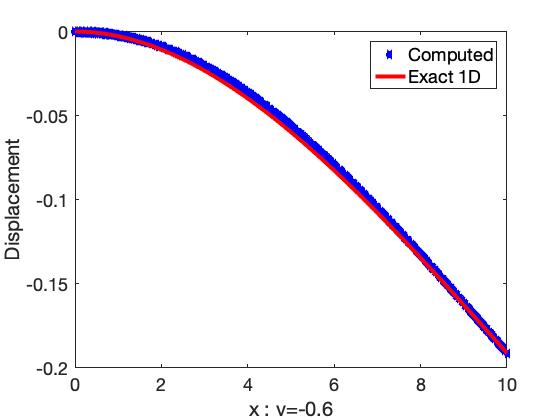}
\includegraphics[width=0.32\textwidth]{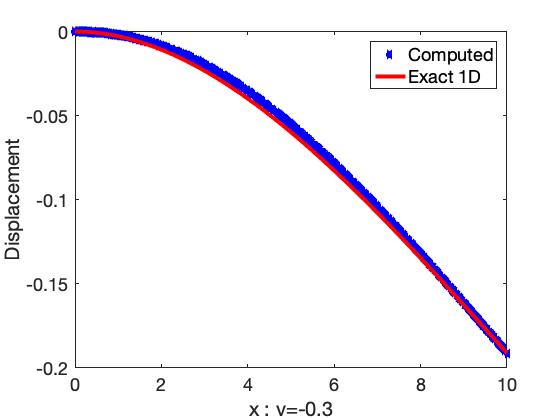}
\includegraphics[width=0.32\textwidth]{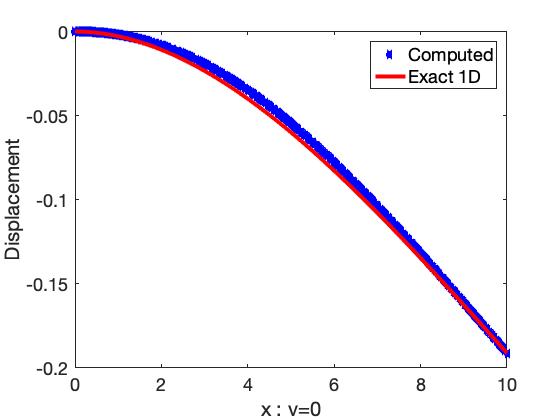}
\includegraphics[width=0.32\textwidth]{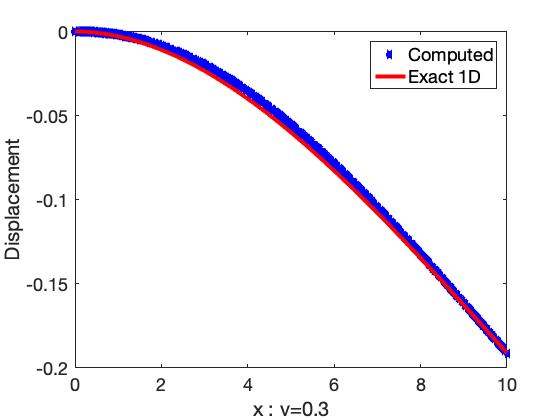}
\caption{Computed deflection of a cantilever beam \label{fig:beam_def}}
\end{figure}

The computed deflection of the cantilever beam is shown in Figure \ref{fig:beam_def} for a few different Poisson's ratios. The deflection is plotted for all $0$-simplices in the primal complex, not just those along the center-line of the beam. Despite a small offset the plots show good agreement with Timoshenko beam theory \cite{timoshenko}.

\subsection{3D Kirsch problem}

\begin{figure}[!t]
\centering
\includegraphics[width=0.5\textwidth]{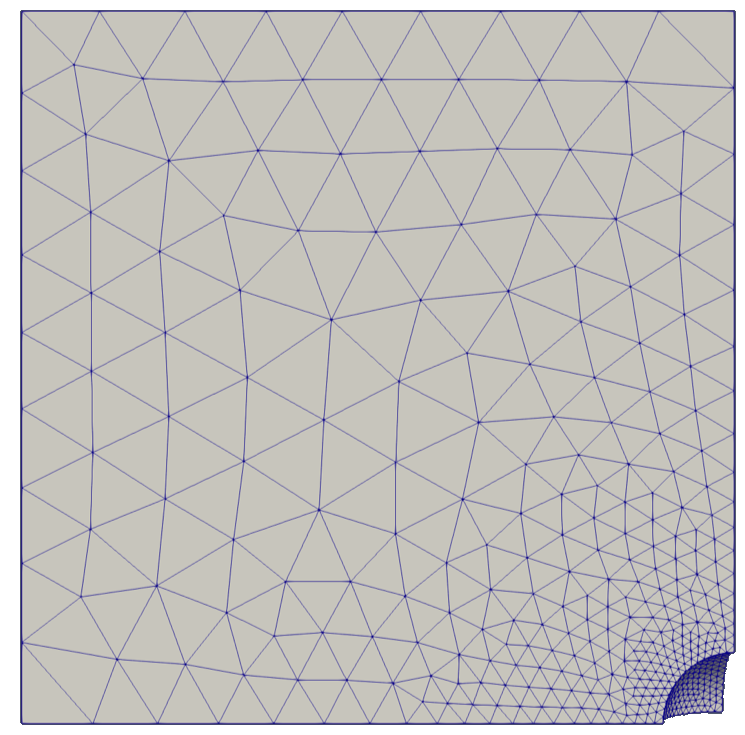}
\caption{Primal complex of one-eighth 3D Kirsch problem \label{fig:kirsch}}
\end{figure}

Finally, the theory is applied to the 3D Kirsch problem: tension applied to a cube with a small spherical hole at the center. The cube has an edge length of $20$ and the spherical hole has a radius of $1$. The present simulation takes advantage of symmetries and models only one eighth of the full geometry, as shown in Figure \ref{fig:kirsch}. The primal complex used has $3,408$ vertices and $21,142$ edges.

\begin{figure}[!t]
\centering
\includegraphics[width=0.32\textwidth]{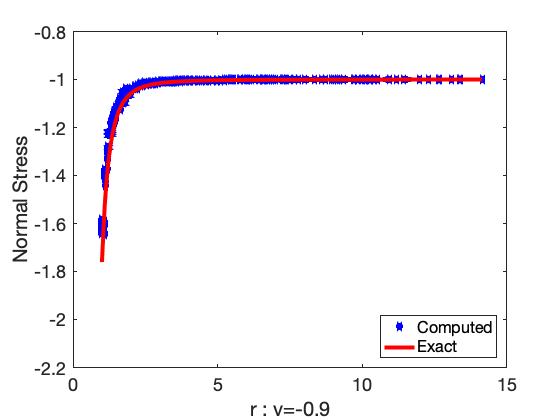}
\includegraphics[width=0.32\textwidth]{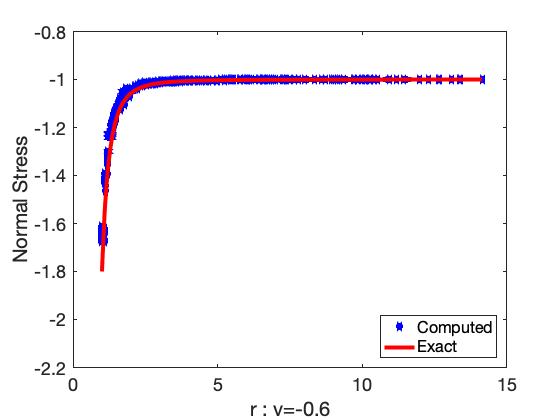}
\includegraphics[width=0.32\textwidth]{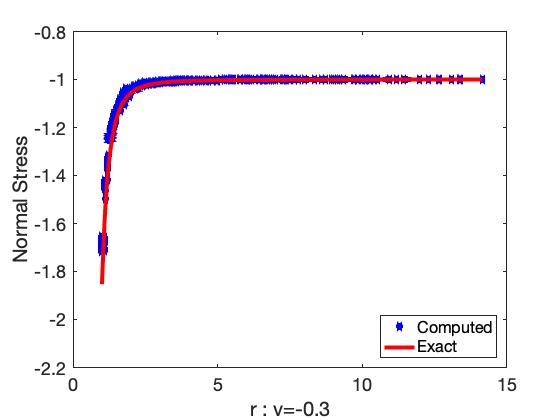}
\includegraphics[width=0.32\textwidth]{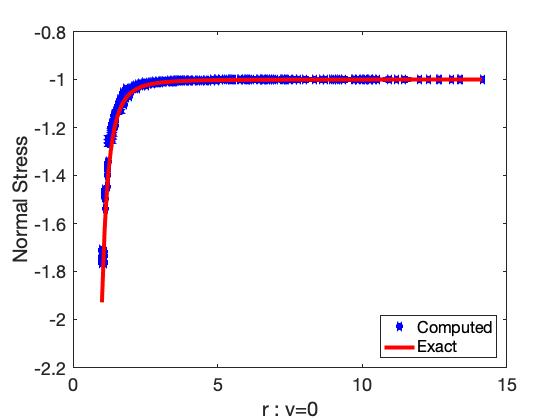}
\includegraphics[width=0.32\textwidth]{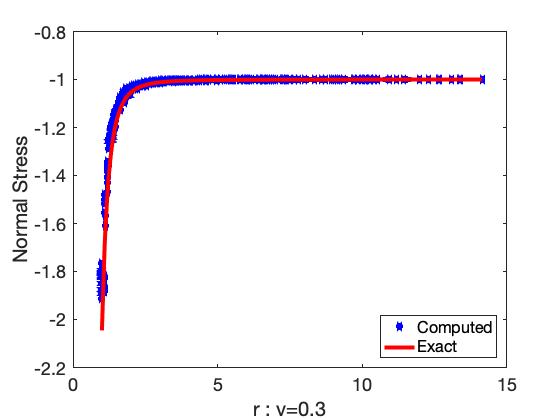}
\caption{Computed normal stress for the 3D Kirsch problem on the cube's mid plane as a function of radial distance from the center of the spherical hole. \label{fig:kirsch_stress}}
\end{figure}

\begin{figure}[!t]
\centering
\includegraphics[width=0.32\textwidth]{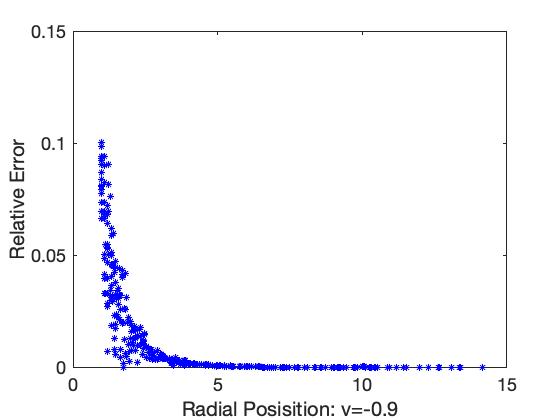}
\includegraphics[width=0.32\textwidth]{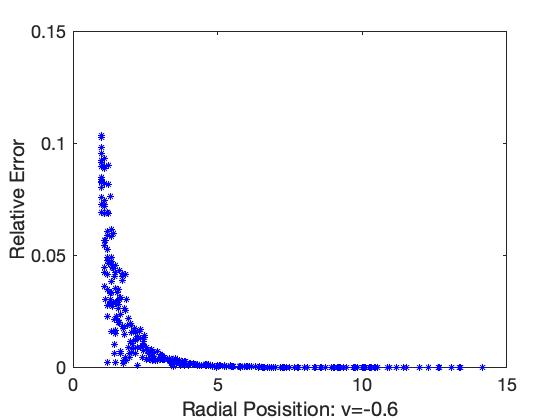}
\includegraphics[width=0.32\textwidth]{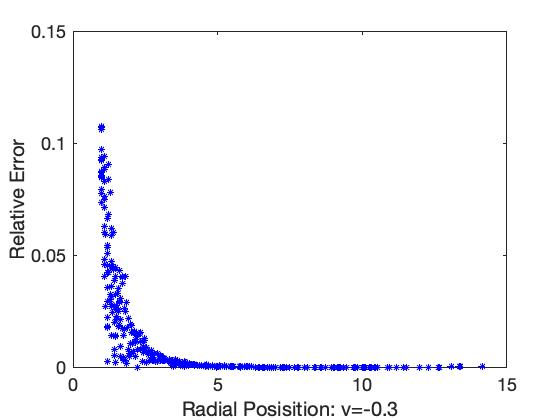}
\includegraphics[width=0.32\textwidth]{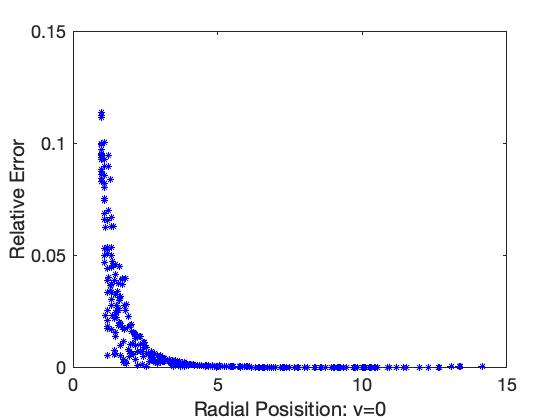}
\includegraphics[width=0.32\textwidth]{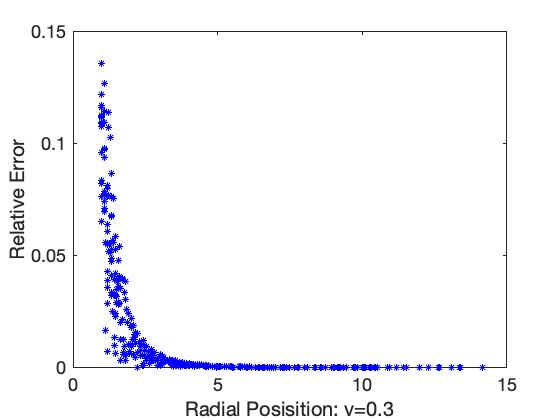}
\caption{Relative error in normal stress for the 3D Kirsch problem on the cube's mid plane as a function of radial distance from the center of the spherical hole. \label{fig:kirsch_err}}
\end{figure}

Figure \ref{fig:kirsch_stress} shows the normal stress on the mid plane of the cube, as a function of radial distance from the center of the spherical hole. While agreement with theory looks reasonably good, the error grows to as high as $\sim\%14$ close to the hole, which is shown in Figure \ref{fig:kirsch_err}. Here the radial gradient in the normal stress is the highest, and most affected by the non-local nature of the discrete musical isomorphisms. This is similar to the results for the cylinder twist. Again, comparing with results obtained on a coarser mesh, the error decreases with increased mesh density, supporting the hypothesis that the discrete musical isomorphisms are the source of the error.

%
\section{Discussion and conclusions}\label{sec:conc}
%

This article presented a geometric description of linear elasticity using discrete exterior calculus. The description uses a vector-valued primal $0$-cochain as the primary unknown in the boundary value problem. The discrete exterior (combinatorial) derivative of this vector-valued $0$-cochain is a vector-valued $1$-cochain representing displacement differences, while the corresponding internal forces form a dual vector-valued $2$-cochain.  Notably, previously proposed formulations \cite{yavari:2008,tonti:2013} take the same view, but without specifying how to prescribe the constitutive relation between the two cochains. In fact, direct relation between the two, that is unique for the entire complex, cannot be established in the general case - a situation similar to irregular lattice structures. Therefore, the following argument is used. 

The material constitutive law is derived from macroscopic experiments and links intensive, i.e. point-wise, quantities: stress and strain. This requires the introduction of a macroscopic constitutive relation at points, in our case at primal $0$-cells. The primal vector-valued $1$-cochain can be considered as a flat of the displacement gradient in continuum mechanics, while the dual vector-valued $2$-cochain can be considered as a flat of the Cauchy stress tensor. Hence, the sharp of the primal vector-valued $1$-cochain approximates the displacement gradient at primal $0$-cells. A prescribed macroscopic constitutive relation can then be applied to the symmetric part of the discrete displacement gradient to find the discrete stress tensor, the flat of which defines the internal forces. This is presented for linear elastic isotropic materials, but there is no reason more complex finite-strain relationships could not be applied. These forces are diverged back to the primal  $0$-cell to balance the linear momentum. A force closure is presented to complete the divergence at the boundary of the complexes, where dual $3$-cells are not fully closed by dual $2$-cells. 

It was shown that the action of applying the constitutive law could be absorbed into the Hodge star. Therefore, the governing equations can be solved as a Laplace equation using a non-local and non-diagonal material Hodge star. Additionally, it was shown that the basic application of the constitutive law can be simplified to minimise the use of approximate musical isomorphisms. One important consequence of this absorption is that the matrix-valued Hodge-stars can be calculated for the primal $1$-cells of any irregular lattice, thus solving the problem of finding a relation between local element properties and macroscopic properties. With the present formulation, the matrix-valued Hodge-stars are applicable to lattices forming irregular tetrahedra, and provide the link between displacement difference and internal force vectors; the only requirement is to construct the circumcentric dual for the additional geometry information required. However, this can be extended readily to more general complexes, for example by starting with a dual vector-valued $0$-cochain as the primary unknown.

Numerical simulations of several classical problems with analytic solutions were presented to validate the formulation. Accurate solutions were obtained throughout the range of Poisson's ratios from $-0.95$ through $0.45$. Errors in the computation of stresses was attributed to the non-local nature of the musical isomorphisms used. This error diminished with mesh refinement, supporting this hypothesis.
  
The proposed DEC formulation of elasticity corresponds to the primal/direct formulation in the finite element method, where the nodal displacements are the primary unknowns. In such case, the displacement field is continuous, while the stress field is discontinuous across element boundaries. This is the situation with the proposed DEC formulation, since the internal forces at dual $2$-cells are calculated as flats of the Cauchy stress tensor, i.e. they are averages of two different vectors arising from the two dual $3$-cells adjacent to each $2$-cell. Work is ongoing on a DEC formulation that corresponds to the mixed formulation in the finite element method, where both displacements and internal forces are unknowns. This will ensure continuity of both displacement field and internal forces across dual $2$-cells.

\section{Funding}

This work was supported by the Engineering and Physical Sciences Research Council [EPSRC Fellowship: Geometric Mechanics of Solids: new analysis of modern engineering materials, EP/N026136/1, 2017-2022].

%
    \bibliographystyle{model1b-num-names}
    \bibliography{thesis}
%

\end{document}